\documentclass[pre,aps,superscriptaddress,floatfix]{revtex4-2} 

\usepackage[english]{babel}
\usepackage{float}
\usepackage{color}
\usepackage{graphicx}
\usepackage{dcolumn}
\usepackage{bm}
\usepackage[normalem]{ulem}
\usepackage{xcolor}
\usepackage{natbib}
\usepackage{amssymb}
\usepackage{bm}
\usepackage{hyperref}

\begin{document}

\title{Statistical domain wall roughness analysis through correlations}

\author{Pablo Domenichini}
\affiliation{Facultad de Ciencias Exactas, Universidad Nacional de Salta.}
\affiliation{CONICET Salta-Jujuy - Instituto de Investigaciones en Energía No Convencional.}
\author{Gabriela Pasquini}
\affiliation{CONICET-Universidad de Buenos Aires, IFIBA, Buenos Aires, Argentina.}
\affiliation{Universidad de Buenos Aires, FCEyN, Departamento de F\'{\i}sica. Buenos Aires, Argentina}
\author{María Gabriela Capeluto}
\affiliation{CONICET-Universidad de Buenos Aires, IFIBA, Buenos Aires, Argentina.}
\affiliation{Universidad de Buenos Aires, FCEyN, Departamento de F\'{\i}sica. Buenos Aires, Argentina}
\affiliation{maga@df.uba.ar}

\date{\today}

\begin{abstract}
The geometry and morphology of magnetic domain walls (DWs) are closely related to their dynamics when driven by external forces. Under some reliable approximations DWs can be considered self-affine interfaces, so universal laws govern their behavior. On the other hand, large-scale DW structure has been less explored so far. Recently, it has been shown that bubble-like magnetic domains can be strongly deformed on a large scale by applying alternating (ac) magnetic field pulses. In the present work, we conduct a comprehensive analysis of DW structure at both small and large length scales in bubble-like domains present in ferromagnetic thin films with perpendicular anisotropy, focusing on its initial evolution under the application of ac magnetic pulses. Results obtained from the widely used roughness correlation function $B(r)$ and its corresponding structure factor, are consistent with those obtained from the spatial autocorrelation function of DW fluctuations. Whereas the roughness exponent slightly increases during the ac evolution, a strong deformation is observed at a large scale, where a striking periodicity (statistically speaking) is observed. This period is probably determined by the boundary conditions and a characteristic intrinsic length. \\
The following article has been accepted at Physical Review B. After it is published, it will be found at  \href{https://journals.aps.org/prb/}{Link}.
\end{abstract}

\pacs{Valid PACS appear here}

\maketitle

\section{ Introduction}\label{sec:Introduction}

The control of magnetic properties in thin films is crucial for the design of devices for technological applications \cite{Art-APL116-Mohammed,Art-PhyRep958-Kumar, Puebla2020}. 

For this purpose, it is key to deeply understand the physics that governs the behavior and characteristics of magnetic domain walls (DWs) in these systems. In particular, it is well known that the geometry and morphology of DWs are closely related to their dynamics when driven by external forces \cite{lemerle1998domain,Art-PRL99-Metaxas,ferre2013universal,Art-PRL113-Gorchon, Art-PRL117-Jeudy}. The basic physical reason for this close relationship is quite intuitive: as the domain interface is driven in a disordered media, some DW's segments remain more pinned than others, and the interface becomes rough. Moreover, under some reliable approximations, both the DW dynamics and morphology display universal laws that govern the behavior of elastic systems in disordered landscapes \cite{lemerle1998domain,Art-PRL99-Metaxas,ferre2013universal,Art-PRL113-Gorchon, Art-PRL117-Jeudy, Art-PRL89-Tybell, Art-EPL87-Doussal, Art-NatC4-Laurson, Art-RMP66-Blatter, Art-RPP80-Reichhardt}.

An emblematic example of this connection is the link between the creep exponent $\mu$, which characterizes the dependence of the effective energy barriers (and therefore the mean DW velocity) on the external drive, and the equilibrium roughness exponent $\zeta_{eq}$, which describes how the roughness of the interface grows as a function of the length scale at zero drive. In fact, it can be shown that the creep exponent for a d-dimensional interface can be expressed in terms of the equilibrium roughness exponent as $\mu = (d + 2 \zeta_{eq} - 2)/(2 - \zeta_{eq})$ \cite{lemerle1998domain, Art-PRL94-Paruch, Art-PRB62-Chauve, Art-PB407-Agoritsas, Art-PRB101-Jordan}. In a one-dimensional DW belonging to the so-called Edwards-Wilkinson universality class, the theoretical equilibrium roughness exponent is $\zeta_{eq} = 2/3$, which should imply a creep exponent $\mu = 1/4$, a value that has been corroborated in many experiments \cite{ferre2013universal, Art-PRL113-Gorchon, Art-EL68-Repain, Art-PRL89-Tybell, Art-PRB101-Jordan, Art-PRL99-Metaxas, Art-APL112-Quinteros, Art-JPD54-Quinteros}. 

On the other hand, the experimental determination of the roughness exponent is much more challenging, leading to a wide spread of reported values found in the literature \cite{Art-PRB104-Burgos, Art-IEEE45-Kang, Art-PRL110-Kyoung, Art-PRB101-Jordan, Art-PRB-104-Albornoz}. The main physical intrinsic reason for this spread is the fact that DWs are generally frizzed in non-equilibrium configurations, so the DW morphology is described by different roughness exponents at different length scales \cite{Art-PRL97-Kolton, Art-PRB79-Kolton, Art-CRP14-Ferrero, Art-PRB-104-Albornoz,Art-PRB104-Burgos}. Recently, Albornoz and coworkers \cite{Art-PRB-104-Albornoz} proposed a method to experimentally obtain the characteristic distances that determine the crossover between the different regimes: the optimal creep length $l_{opt}$, below which the interface may be considered in equilibrium, and the avalanche length $l_{av}$, associated with the average size of avalanches from which the disembedding processes is described. This method is explained in more detail in Section \ref{sec:ResultsAndDis}-B. On the other hand, the main technical reason that makes the exponent measurement difficult is the practical determination of the range where theoretical approximations are valid, i.e. the range of lengths where criticality holds \cite{Art-PRB-104-Albornoz}. 

In fact, the interface width that characterizes the mean size of fluctuations  in a DW segment of length $r$,  $ w_r = \sqrt{ {\langle (u(z) - \langle u\rangle)^2 \rangle}_r} $, and the corresponding roughness correlation function, $B(r) = \overline{<[u(r+z) - u(z)]^2>}$, are expected to scale as $w_r ^2 \propto B(r) \propto r^{2 \zeta}$ for a self-affine interface \cite{Libro-Barabasi-fract}. In both cases, $z$ is the coordinate along the main interface direction, and $u(z)$ is the relative displacement in the perpendicular direction. In real interfaces, a necessary condition to fulfill self-affinity is $l_{res}<r \ll L$, where $l_{res}$ is the experimental resolution and $L$ is the total DW length.

In this framework, an issue that has not been much explored so far is the role that plays the boundary conditions in the characteristic of large-scale DW deformations, and whether these patterns affect or not the short-scale DW fluctuations, where criticality is expected to hold. Interestingly, recent magneto-optical experiments carried out in ultra-thin ferromagnetic films with perpendicular magnetic anisotropy (PMA), have shown that large-scale DW deformations are dramatically enhanced under the applications of alternating (ac) magnetic field pulses \cite{Art-PRB99-Domenichini, Art-PRB103-Domenichini}, while no so significant changes are observed in the roughness exponent.

In this work, we perform a detailed analysis of the roughness correlations at short and large spatial scales of bubble-like domain walls in ferromagnetic thin films with PMA, and their evolution under the applications of ac pulses. We also obtain rich information about this evolution at different spatial frequency scales by analyzing the autocorrelation of DW fluctuations. On the one hand, as expected, the high-frequency fluctuations follow a universal power law with the same exponent holding for $B(r)$ in the short distances range. On the other hand, we show that a large-scale periodic modulation of the interface is established, whose amplitude grows with an increasing number of ac pulses. Moreover, the periodicity of this modulation is on the order of the characteristic avalanche length, obtained from the analysis of the effective roughness exponent. The underlying physics beyond these periodic structures is discussed. 

\section{Experimental}\label{sec:Experimental and methods} 

Images of magnetic domains were obtained using a homemade polar Kerr effect microscope described in detail in Ref. \cite{Art-PRB99-Domenichini}. With this setup, images with 10$\times$ magnification and $1\, \mu$m spatial resolution can be obtained. Specially designed Helmholtz coils allow applying squared magnetic pulses with a maximum amplitude of 700 Oe and a minimum duration of $\tau =1$ ms, in the direction normal to the sample. Magnetic domains were grown on two reflective ultra-thin ferromagnetic films with PMA: a Pt$ (4 \, $nm$)/$Co$(1 \, $nm$)/$Pt$(8 \, $nm$)$ monolayer (S1) \cite{Art-PRL99-Metaxas}, and a Al$\, (5\, $nm$) [$Co$ \, (0.2\, $nm$)/$Ni$ \, (0.8\, $nm$)]_4 / $Pt$ \, (6$\,nm$)$ multilayer (S2) \cite{Art-APL108-Rojas}.
 
\begin{figure}[h]
  \centering
    \includegraphics[scale=0.6]{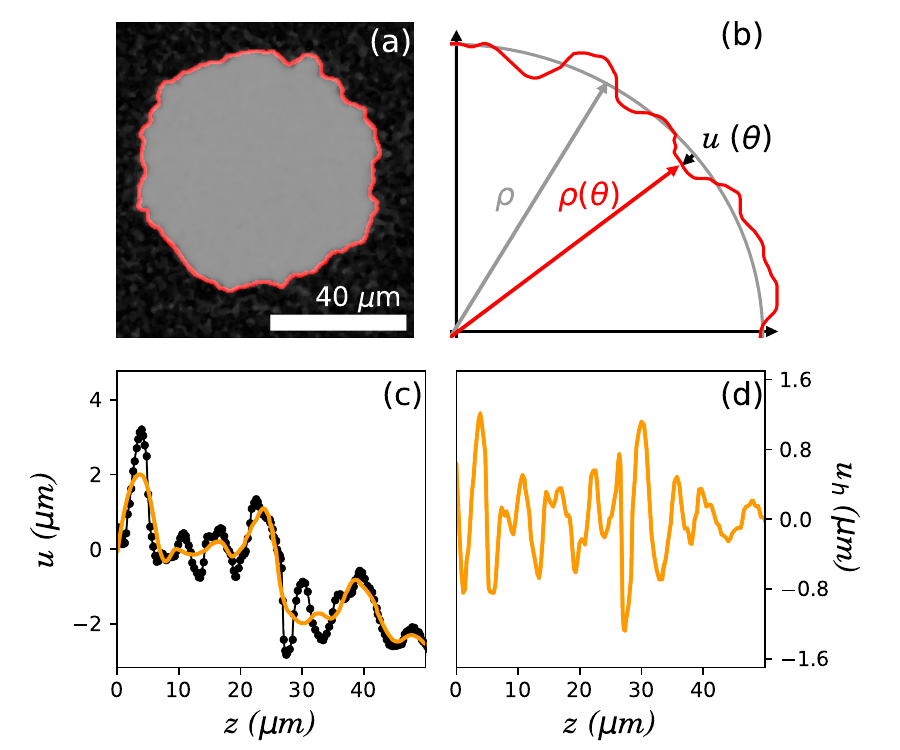}
    \caption{(a) Image of a domain and its contour (red line) obtained after a nonlinear thresholding process. (b) Domain profile and polar parametrization where $\rho (\theta)=\rho + u(\theta)$. (c) local DW relative displacement $u(z)$ in Cartesian coordinates (black dots), where $z= \rho \theta$. The orange line represents the fluctuations in DW displacement $u_l^r(z)$ after applying a moving average smoothing filter with a window size of $r = 6 \mu m$. (d) Corresponding high-frequency profile obtained as $u_h^r(z)= u(z)-u_l^r(z)$.}
    \label{fig:Fig-Perf}
\end{figure}

The protocol (pulse sequence) used to study the ac dynamics was established in a previous work \cite{Art-PRB99-Domenichini} and it is briefly described here. After nucleation, domains are grown up to an area of around 4700 $\mu $m$ ^2$ by applying magnetic field pulses, between 100 - 250 Oe, and periods between 1 - 50 ms. Then a sequence of $N$ square ac pulses of null mean, with amplitude $H$ and period $\tau$ is applied. The magnitudes of $H$ and $\tau$ are chosen in a way to generate a DW displacement of 6 $\mu$m during each half-period. Images of the domain evolution are acquired after applying each pulse and further processed to remove the background and enhance their contrast. Finally, after applying a threshold filter, images are binarized in order to obtain the domain's contour, i.e. the boundary between regions of different magnetization (See Ref.\cite{Art-PRB99-Domenichini} for detailed descriptions). An example of an already binarized image and its contour is shown in Figure \ref{fig:Fig-Perf}(a).

The basic DW dynamics characterization consists of determining the relationship between the mean DW velocity $v$ and the external drive; in our case, the applied magnetic field $H$. The method used to measure the $v(H)$ dependence essentially consists of applying successive magnetic field pulses of positive (negative) direction (relative to the nucleation field direction) and tracking the advance (recoil) of the DW. The mean DW velocity can be estimated from the ratio between the average DW displacement and the pulse width (For more details see Ref. \cite{Art-PRB99-Domenichini}). 

Regarding the morphological characterization, in practical terms, the first step in determining roughness is to obtain a parametrization of the domain interface. Since our experimental configuration allows obtaining images of the whole domain, which has an approximately circular shape, a polar parametrization of the DW is performed; as shown in Figure \ref{fig:Fig-Perf}(b), by using a system centered at the domain's centroid, each DW point is specified by an angle $\theta$ and a radius $\rho(\theta)=\rho + u(\theta)$, where $\rho$ is the average radius and $u(\theta)$ is, therefore, the local DW displacement. The DW contour is then mapped in Cartesian coordinates using a linear transformation $z=\rho \theta$ and $u(z)=u(\theta)$, as shown in Figure \ref{fig:Fig-Perf}(c) (black dots). Evidently, to be able to carry out the above procedure, $u$ needs to be a single-valued function of $\theta$, a condition that is not fulfilled in strongly distorted domains. To avoid this problem, all the analysis of the ac evolution presented in this work is conducted within the first 30 ac cycles when, in all cases, DWs are single-valued curves.

Given the statistical nature of DW dynamics and in order to ensure reliable results, all the reported observables were estimated by averaging the results obtained from ten experimental realizations conducted for each of the reported conditions.

\section{\label{sec:ResultsAndDis} Results and Discussion}

In this section, we first describe how we define roughness at different spatial frequencies. We then obtain the corresponding roughness correlation function $B(r)$ and the associated structure factor $S(q)$. Next, we introduce the autocorrelation function $A(\ell)$ as a tool to retrieve the key features that describe the DW evolution. Specifically, as a method to analyze the different spatial frequency scales of DW fluctuations and unveil hidden periodic structures in the DW fluctuations as well.

\subsection{\label{sec:CrossLen} DW roughness at different length scales}

DW roughness characterization has been done after having applied field pulses in the creep regime range, where the velocity dependence on the magnetic field $H$ and temperature $T$ is predicted to be

\begin{equation}
   v(H,T)= v_d \,\, exp\left ( \frac{T_d}{T} \left [1- \left ( \frac{H}{H_d} \right ) ^{-1/4} \right ] \right ),
    \label{eq:vh}
\end{equation}
where $H_d$ and $T_d$ are the depinning magnetic field and temperature respectively and $v_d$ is the velocity for $H=H_d$. In fact, as reported in previous work \cite{Art-PRB99-Domenichini}, a linear relationship between $ln(v)$ and $H^{-1/4}$ is observed in a broad magnetic field range for both samples, confirming that domains are in the creep regime with the expected dynamic exponent $\mu = 1/4$. 

A first observable that is typically used to quantify 
roughness is the mean square roughness $\langle (u-\langle u\rangle)^2\rangle$, whose definition coincides with the square of the mean width of the whole interface, $w_L^2$, as long as $\langle u \rangle$ = 0 and averages are computed over the entire DW length $L$. 

By simply inspecting the amplitude of the DW fluctuations in Figure \ref{fig:Fig-Perf}(c), it is evident that they occur at two distinct scales. A low-frequency scale of fluctuations characterized by large DW deformations that modulate higher-frequency fluctuations. The latter is expected to be well-described by a power-law behavior in the roughness correlation function. As mentioned before, we are interested in studying how the DW fluctuations behave at these different length scales or spatial frequencies. With this scope, a moving average smoothing filter (MASF) of window size $r$ is applied along the profile $u(z)$, in a way that fluctuations that vary faster than $1/r$ are averaged, and only the low-frequency scale fluctuations remain in the averaged signal $u _{l}^{r}(z)$ \cite{Libro-Filters-Smith}. In a continuous interface, a possible implementation of the MASF to calculate $u_{l}^{r}(u)$ is 

\begin{equation}
    u_{l}^{r}(u)= \frac{1}{r} \int ^{z + \frac{r}{2}}_ {z - \frac{r}{2}}u(z^{'}) d z^{'} ,
    \label{eq:ularge}
\end{equation}
from this expression, implementing a discrete interface becomes straightforward. An example for $r=6 \, \mu m$ is shown in Figure \ref{fig:Fig-Perf}(c) (orange line). Accordingly, the profile that characterizes the high-frequency scale DW fluctuations $ u_{h}^{r}(z)$ (Figure \ref{fig:Fig-Perf}(d)) is simply obtained by the subtraction of $ u_{l}^{r}(z)$ from $u(z)$:

\begin{equation}
    u_{h}^{r}(z)= u(z)-u_{l}^{r}(z) .
    \label{eq:usmall}
\end{equation}

\subsection{\label{sec:RougFun} Roughness correlation function $B(r)$ and effective exponent}

Among the observables that allow analyzing the DW roughness, one of the most used in the literature is the roughness correlation function 

\begin{equation}
    B(r) = \overline{<[u(r+z) - u(z)] ^2> _L}.
    \label{eq:Rug}
\end{equation}
where the subscript $L$ indicates that $z$ values are averaged over the total length of the interface and the top bar indicates the statistical average. $B(r)$ therefore quantifies the correlation between relative displacements $u$ of interface points that are a distance $r$ apart. Specific numerical expressions used in this work are discussed in Appendix A.

Figure \ref{fig:Fig-BS}(a) shows examples of $B(r)$ obtained from individual measurements (curves in gray levels) for 10 domains grown on S2 from the same nucleation center up to a mean radius $\rho \approx 40 \,\mu$m with $H=150 \, Oe$, together with the average $B(r)$ curve (red line).

As mentioned in the introduction, for a self-affine interface the roughness correlation function is expected to follow a power law \cite{Art-JPCM33-Bustingorry, lemerle1998domain, Art-PRB-104-Albornoz}

\begin{equation}
    B(r) \sim B_o \left ( \frac{r}{r_o} \right ) ^{2 \zeta},
    \label{eq:Rug-exponent}
\end{equation}
where the prefactor $B_o$ characterizes the typical fluctuation size in a DW segment of length $r$, expressed in units of $r_o$, and $\zeta$ is the effective roughness exponent which characterizes the power law scaling. Therefore, from Equation \ref{eq:Rug-exponent}, the roughness exponent can be determined from the slope of the linear fit on a logarithmic scale in the range where the scaling holds (green line in Figure \ref{fig:Fig-BS}(a)). 

In a one-dimensional self-affine interface, the theoretical equilibrium roughness exponent is predicted to be $\zeta_{eq} = 2/3$. However, real magnetic DW interfaces are not in equilibrium conditions, and an effective $\zeta$ (ranging between 0.6 and 0.8 can be obtained \cite{Art-PRB99-Domenichini, Art-PRB104-Burgos, Art-IEEE45-Kang, Art-PRL110-Kyoung}) depending on parameters such as field and temperature, as well as the particular procedure used in its estimation. 
There are two main reasons for this spread: on the one hand, the experimental distance range generally covers more than one dynamic regime (therefore several exponents coexists); on the other hand, self-affinity would be a good approximation only in a particular distance range. As a consequence, there is a limited range of $r$ where the power law holds, whose delimitation is not obvious. As a general criterion in this work, linear fits were performed within the interval of distances determined by the optical resolution (1 $\mu$m) on the left side and, on the right side, by the value of $r$ that results in a Pearson correlation coefficient of $R^2>0.995$. In the particular case of the curves shown in Figure \ref{fig:Fig-BS}(a), this procedure resulted in $\zeta = (0.65 \pm 0.03)$. 

\begin{figure}[H]
  \centering
    \includegraphics[scale=0.75]{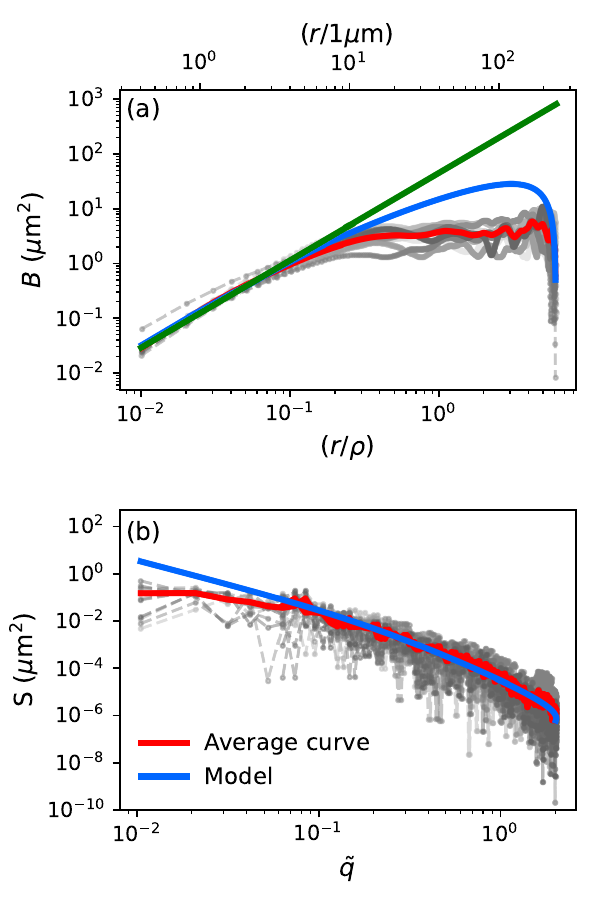}
    \caption{Roughness correlation function (a) and structure factor (b) obtained from 10 realizations from the same nucleation point in the sample S2 (gray lines) and the average curves (red line). The linear fit on $B(r)$ used to obtain $\zeta$ is shown in green. Curves obtained from the $S(q)$ model with the fitting parameters are shown in blue (see text).}
    \label{fig:Fig-BS}
\end{figure}

The procedures described above were applied to obtain the mean square roughness $\langle u^2\rangle$ and the corresponding $B(r)$ during the ac evolution. Figure \ref{fig:lav}(a) and (b) show the overlapping of magnetic domains in a typical evolution during the application of ac pulses (up to $N=30$) with amplitude of $130 \, Oe$ for S1, and $180 \, Oe$ for S2. It can be seen with the naked eye that DW roughness increases during evolution. Consistently, as the number of pulses increases, the mean square roughness $\langle u^2 \rangle$ and its dispersion increase (Figure \ref{fig:lav}(c) and (d)). From the linear fit of the mean $B(r)$ curves, and following the criterion described above, the evolution of the roughness exponent with the number of applied ac pulses (N) was obtained. As it is shown in Figure \ref{fig:lav}(e) and (f), the roughness exponent slightly grows with the number of pulses on both samples. It can be also seen that none of these results show significant differences when varying the magnetic field amplitude in the range used in our experiments. 

\subsection{\label{sec:StrucFact} Structure factor $S(q)$ and avalanche length}

The roughness exponent $\zeta$ is also related to the power spectrum in the reciprocal space or structure factor 

\begin{equation}
   S(q) = \overline{\tilde{u} (q) \tilde{u} ^* (q)},
   \label{eq:Sq}
\end{equation}

\noindent
where $q = 2 \pi /u$ is the reciprocal space coordinate, and $\tilde{u} (q)$ is the Fourier transform of $u(z)$.

\begin{figure}[h]
  \centering
    \includegraphics[scale=0.49]{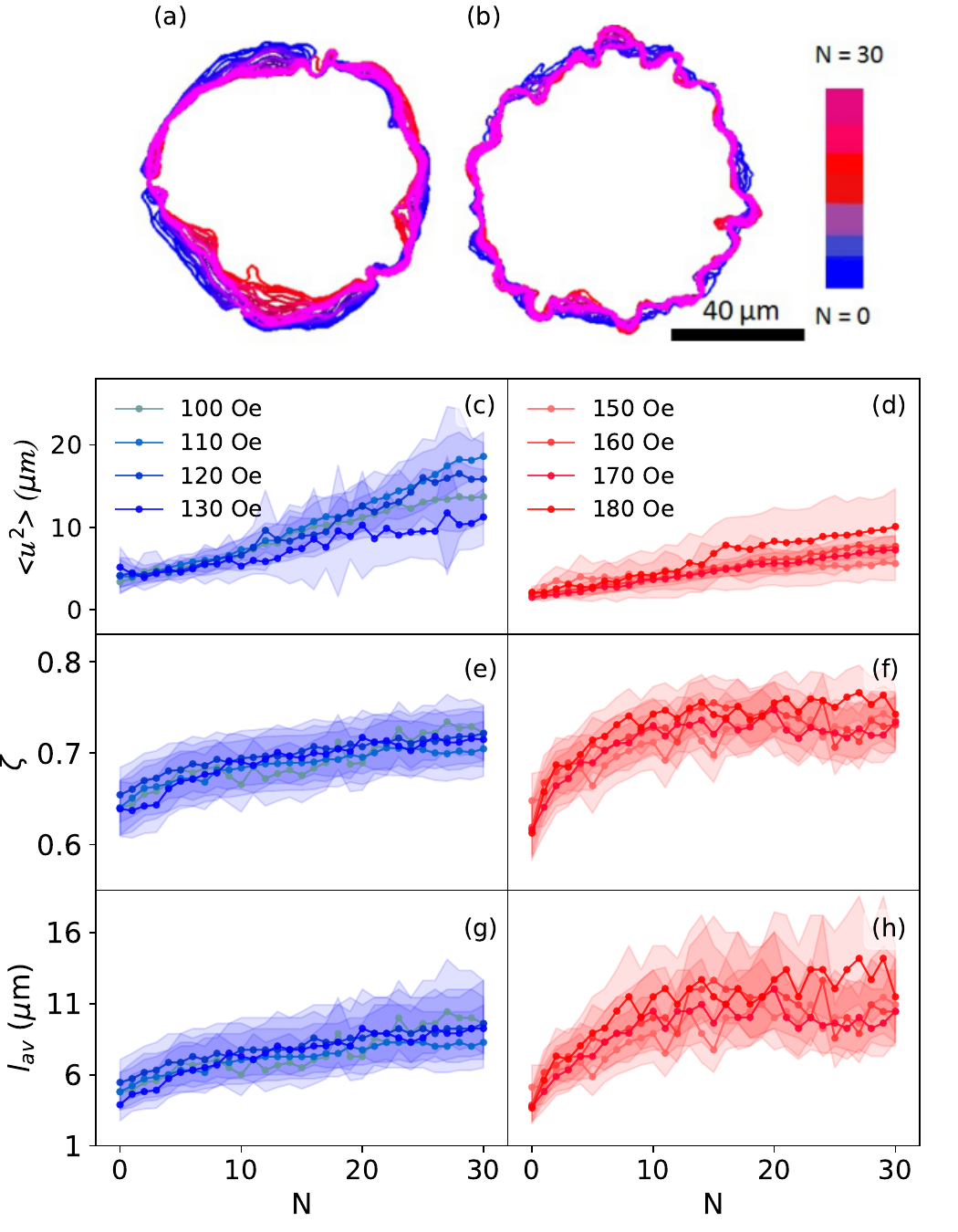}
    \caption{Upper panels show the overlapping of domain contours evolving as the number of applied ac pulses increases, from $N = 0$ (blue) to $N = 30$ (magenta), in samples S1 (a) and S2 (b). The following panels show the corresponding evolution of the mean square roughness $ \langle u ^2\rangle$ (c and d), roughness exponent $\zeta$ (e and f), and avalanche length $l _{av}$ (g an h) as a function of $N$ for different field amplitudes $H$. In all panels, shaded areas indicate the dispersion of each observable. } 
    \label{fig:lav}
\end{figure}

For discrete interfaces, a discrete expression for $S(q)$ in terms of $u(z)$ can be obtained by replacing the discrete Fourier transform of $u(z)$, $\tilde{u} (q_n)=(1/N_p)\sum_{j}^{N_p} u_j exp(-i q_n u_j)$, in Equation \ref{eq:Sq}, to obtain \cite{Libro-FeroMag-Guyonnet}, for each individual curve:

\begin{equation}
    S(q_n) =   \frac{1}{N_p^2} \Bigg |  \sum_{j=1} ^{N_p} u_j exp(-iq_n u_j) \Bigg|^2,
    \label{eq:Su}
\end{equation}

\noindent  where $N_p$ is the number of points in the discretization of $u(z)$. 

As it is shown in Refs. \cite{Art-PRB-104-Albornoz, Libro-FeroMag-Guyonnet}, in the range of distances $r << L$, a discrete expression of $B(r)$ can be obtained in term of $S(q_n)$ as
\begin{equation}
B(r_k)=4\sum_{n=1}^{\frac{N_P}{2}-1}S(q_n)[1-cos(q_n r_k)]
    \label{eq:BS}
\end{equation}
with $r_k=k \, \delta r$ and $\delta r$ is the distance between consecutive points in the interface (See Appendix \ref{AppendBrSq} for more details). 

The structure factor is then expected to present a region of $q$ where a power law is fulfilled. In the case where a single $\zeta$ holds at all the involved length scales,
\begin{equation}
    S(q) \approx S _0 \left ( \frac{q}{q _0} \right ) ^{- (1+2 \zeta)},
    \label{eq:Sq-ln}
\end{equation}
where $S_0$ is a constant, and $q_0$ is a scale factor. This scaling would hold for large $q$ values, as long as the range of $q$ allows the interface to be considered continuous (i.e. $1/q >> \delta r$). For discrete interfaces, the scaling law can still be valid, when replacing the coordinate $q/{q _0}$  by \cite{Art-PRB79-Kolton, Art-JPCM33-Bustingorry}:
\begin{equation}
     \tilde{q}= 2 \sin {\frac{q}{2{q _0}}}
    \label{eq:qm}
\end{equation}
 
Figure \ref{fig:Fig-BS}(b) shows $S(\tilde{q})$ computed using Equations \ref{eq:Su} and \ref{eq:qm} with $q_0$ chosen as $1/ \delta r \approx 2.5 \, \mu$m$^{-1}$,  and the same data-set of Figure \ref{fig:Fig-BS}(a). As observed, it is not possible to fit $S(\tilde{q})$ with a single linear function across the entire $\tilde{q}$ range. In fact, in the general case of a DW elastic interface, a critical exponent for each stationary regime (equilibrium, depinning, or flow) can be predicted. For an Edwards-Wilkinson interface with random pinning disorder, the theoretical values of the roughness exponents for these stationary regimes are $\zeta_{eq}=2/3$, $\zeta_{dep}=1.25$ and $\zeta_{th}=1/2$ respectively. As discussed in the Introduction, Ferrero et al proposed that the different stationary regime prevails in different length scales that are delimited by crossover lengths $l_{opt}$ and $l_{av}$ \cite{Art-CRP14-Ferrero, Art-ARCMP12-Ferrero}. Therefore, the effective measured roughness exponent would be determined by a superposition of all the stationary regimes, each of them dominating in a particular length scale. In this context, a generalization for Equation \ref{eq:Sq-ln} was proposed in Ref. \cite{Art-PRB-104-Albornoz}, which takes different roughness exponent for the different length scales as
 
\begin{equation}
    S(q)= \left ( \frac{1}{S_{eq}(q)+S_{dep}(q)} +\frac{1}{S_{th}(q)} \right ) ^{-1},
    \label{eq:modeloS}
\end{equation}
where

\begin{equation}
    S_{j}(q)=S_o \left ( \frac{q}{q_{opt}} \right )^{-(1+2 \zeta_j )}
    \label{eq:Sj}
\end{equation}
with $q_{opt} = 2 \pi / l_{opt}$ and $j=dep$ or $eq$, and 

\begin{equation}
    S_{th} (q) = S_o \left (\frac{q_{av}}{q_{opt}} \right )^ {-(1 + 2 \zeta_{dep} )} \left (\frac{q}{q_{av}} \right )^ {-(1 + 2 \zeta_{th} )}
    \label{eq:Sth}
\end{equation}
with $q_{av}=2 \pi /l_{av}$ and $q_{opt}=2 \pi /l_{opt}$.

This Expression could be used in principle to fit experimental $S(q)$ data as those shown in Figure \ref{fig:Fig-BS}(b), in order to obtain the parameters $S_{o}$, $q_{av}$ and $q_{opt}$. Basically, in this procedure, $l_{opt}$ was estimated from parameters obtained from experimental $v(H)$ curves (depinning magnetic field and temperature), whereas $l_{av}$ was obtained from the proposed model and the effective roughness exponent. A detailed discussion of this issue is given in Ref. \cite{Art-PRB-104-Albornoz}, and the particular implementation done in this work is detailed in Appendix \ref{AppendModelSq}.

In the range of fields used in this work, $l_{opt-S1} \simeq 0.30 \pm 0.01 \, \mu$m and $l_{opt-S2} \simeq (0.24 \pm 0.01) \mu$m. Figure \ref{fig:Fig-BS} shows, as an example, results obtained from this model in blue lines for $B(r)$ (panel a) and $S(\tilde{q})$ (panel b)  (see Appendix \ref{AppendBrSq} for numerical details). The parameter values found in this particular example were $q_{av}=(1.2 \pm 0.1) \mu $m$^{-1}$ and $S_o= (7.1 \pm 0.2)\times 10^{-8} \mu$m$^2$. As it can be seen in Figure \ref{fig:Fig-BS} the model fits the data for $\tilde{q}\gtrsim 0.08$, which is equivalent to $r\lesssim 30 \mu$m. Notice that, as expected, the range of distances where this model is valid is larger than that where a linear fit can be performed on the log plot of $B(r)$. In fact, the model reproduces the loss of linearity and the downward curvature observed in the logarithmic plot of the $B(r)$ function. However, the model fails to fit the experimental data in the range of large distances (i.e. on the order of the domain size).

Following the same procedure, the evolution of the avalanche length $l_{av}$ with the number of applied ac pulses (N) was obtained for both samples. Figure \ref{fig:lav}(g) and (h) shows this evolution for samples S1 and S2 respectively. According to what was observed in the effective roughness exponent, $l_{av}$ slightly grows with $N$ reaching values between 5 $\mu m$ and 10 $\mu m$. 

Therefore, under the proposed model, the small evolution of the effective roughness exponent is ascribed to a subtle increase in the avalanche length, that increments the weight of the (larger) critical depinning exponent in the effective measured $\zeta$. 

\subsection{\label{sec:Autocorr} Autocorrelation function}

In the previous sections, we deeply discussed the power laws holding in both $B(r)$ and $S(q)$ for $r<<L$ (i.e. high spatial frequency scale). For larger values of $r$ (lower frequencies), these laws are no longer valid due to the existence of large-scale correlations and finite-size effects. Here we introduce the autocorrelation function as an additional tool to analyze the ﬂuctuation in different spatial frequency scales. The autocorrelation function of a signal \cite{Libro-Autocor-Broersen} quantifies the degree of similarity or coherence of a signal to the delayed version of itself, as a function of the delay. It can be either defined for spatial or temporal coordinates. For a continuous 1D interface, the autocorrelation in space for $u(z)$ defined as

\begin{equation}
    A(\ell) =\langle u (z+\ell) u(z) \rangle= \frac{1}{L-\ell} \int ^{L-\ell} _{0} u(z +\ell)u(z)^* d z ,
    \label{eq:Autocorr}
\end{equation}
will quantify the correlation of the fluctuations (o normal displacements)  between interface points separated by distances $\ell$. In Equation \ref{eq:Autocorr}, $u(z)^*$ is the complex conjugated of $u(z)$, and can be replaced by $u(z)$ for real signals, which is the case for DW displacements described above. It can be easily seen that there is a close relationship between this expression and the definition of $B(r)$ in equation \ref{eq:Rug}; the implication of this similarity will be deeply discussed throughout this section. The autocorrelation is a symmetric function and its maximum $A(\ell=0)=\langle u(z)^2 \rangle$ is the mean roughness of the interface. In the present case, this observable has been already directly estimated, and its evolution when applying ac pulses has been presented in Figure \ref{fig:lav}(b). 

By using the linear transformation in the variables $z=\rho \theta$, it is possible to write the autocorrelation in angular coordinates $A(\omega)$ with $\omega=\ell/\rho$, which makes it easier to compare results between domains of different radius. 

\begin{figure}[h]
  \centering
    \includegraphics[scale=0.5]{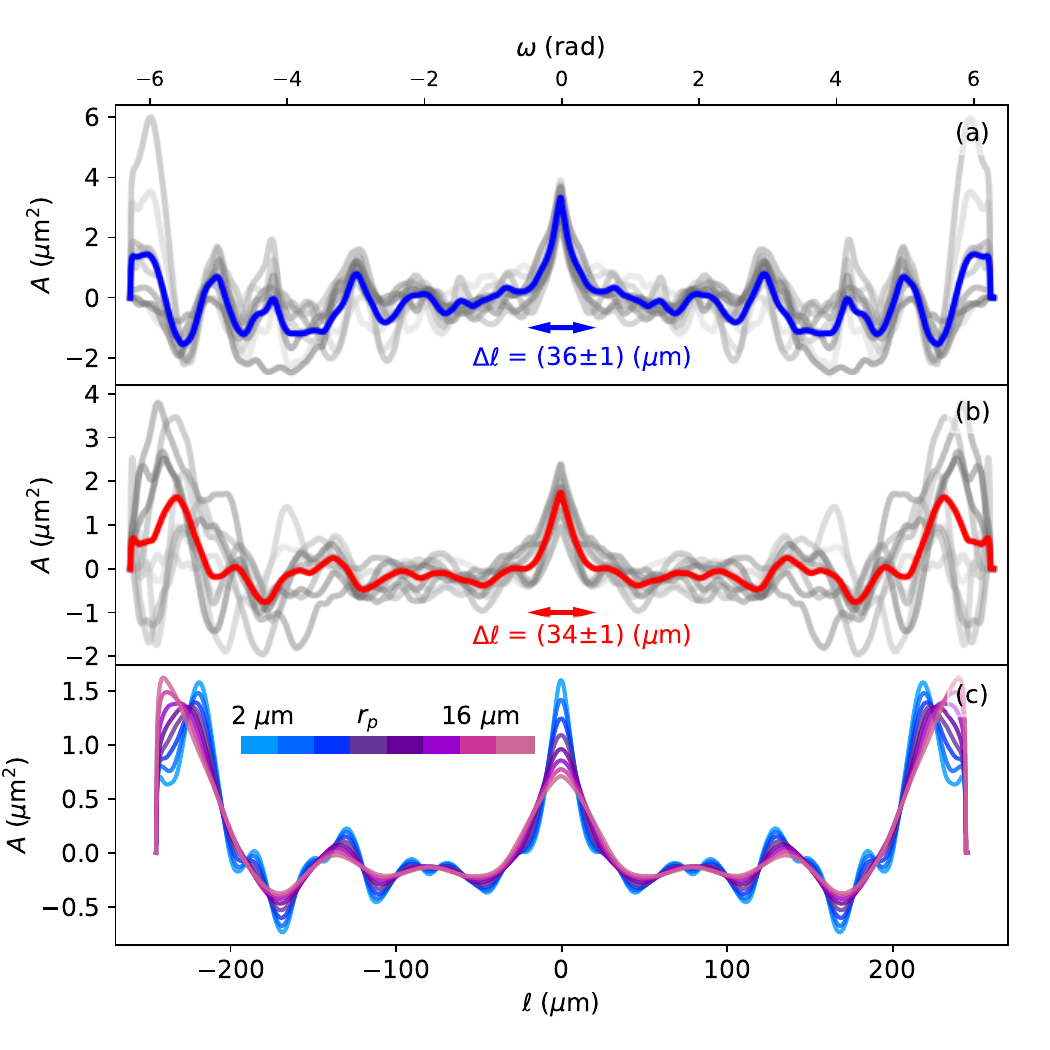}
    \caption{Examples of the autocorrelation function for the samples S1 (a) and S2 (b), where the horizontal arrows show the width of the mean peak. Secondary maxima can be observed in both samples. (c) Autocorrelation of profiles after averaging windows of size $r_p$ between 2 and 16 $\mu$m, obtained from sample S2 with N=0.
    Angular and linear axis are shown for clarity. }
    \label{fig:Fun-Autocorr}
\end{figure}

Panels (a) and (b) in Figure \ref{fig:Fun-Autocorr} show in gray lines examples of the autocorrelation function obtained for domains grown in identical conditions, in samples S1 and S2 respectively. Subtle and small morphological differences observed in the domain shapes obtained in the various realizations are reflected in significant changes in $A(\ell)$. Therefore, in order to obtain more general results, the average of the autocorrelation function $\bar{A}(\ell)$ over 10 experimental realizations is computed (blue and red lines in  the Figures). Both, single measurements and average function, show a main correlation peak and, beyond it, a structured autocorrelation with many maxima and minima. The width  of the main correlation peak serves as an estimator for the correlation length $\Delta \ell$, which represents the maximum distance at which $u(z)$ is similar to the displaced version of itself. For the particular case of measurements presented in Figure \ref{fig:Fun-Autocorr}, and taking $\Delta \ell$ as the width of the central peak, measured between the two closest minima as an estimator, the correlation length results to be $\Delta \ell \sim 35 \mu$m for both samples. This distance being a substantial fraction of the entire domain size suggests that it represents a correlation length associated with low-frequency fluctuations.

In fact, a simple inspection of panels (a) and (b) in Figure \ref{fig:Fun-Autocorr} reveals some other interesting features characterizing the correlations, that are general for all the domains: Beyond the main central peak, other well-defined peaks can be clearly observed, which may be evidence of a periodicity on $u(z)$. The periodicity, estimated as the average distance between the secondary peaks, is a significant fraction of the whole DW length $L$.

To further understand the effects that occur on the fluctuations at different spatial frequency scales, we analyze the average autocorrelation function for $ u_{l}^{r}(z)$ and $ u_{h}^{r}(z)$, for different average window sizes $r$, that in a discrete version are expressed as $r_P= P \delta r$, with $P$ an integer number. As it was mentioned before, $B(r)$ follows a power law for $r<<L$ but, for larger values of $r$, this law is no longer valid. However, $B(r)$ might also contains information about large-scale deformations, linked with the size of the interface. Therefore a deep connection between $B(r)$ and $A(\ell)$ is expected at both  large and small-scale deformation, as will be developed in the following subsections

\subsubsection{Large scale fluctuations}

Panel (c) in Figure \ref{fig:Fun-Autocorr} shows examples of $\bar{A}(\ell)$ for domains grown in S2 using windows ranging from $P = 5$ to $40$, corresponding to distances $r _P$ between $2$ and $16 \, \mu$m. As is expected, the autocorrelation of $u_{l}^{r}(z)$ preserves some of the main features of the autocorrelation of $u(z)$, for example,  the width of the main peak and the presence and positions of the principal and secondary maxima. On the other hand, the height of the principal maximum decreases as the window size increases because the high-frequency fluctuations, that are being blurred by the MASF, have a strong contribution to the central peak.

\begin{figure}[h]
  \centering
    \includegraphics[scale=0.5]{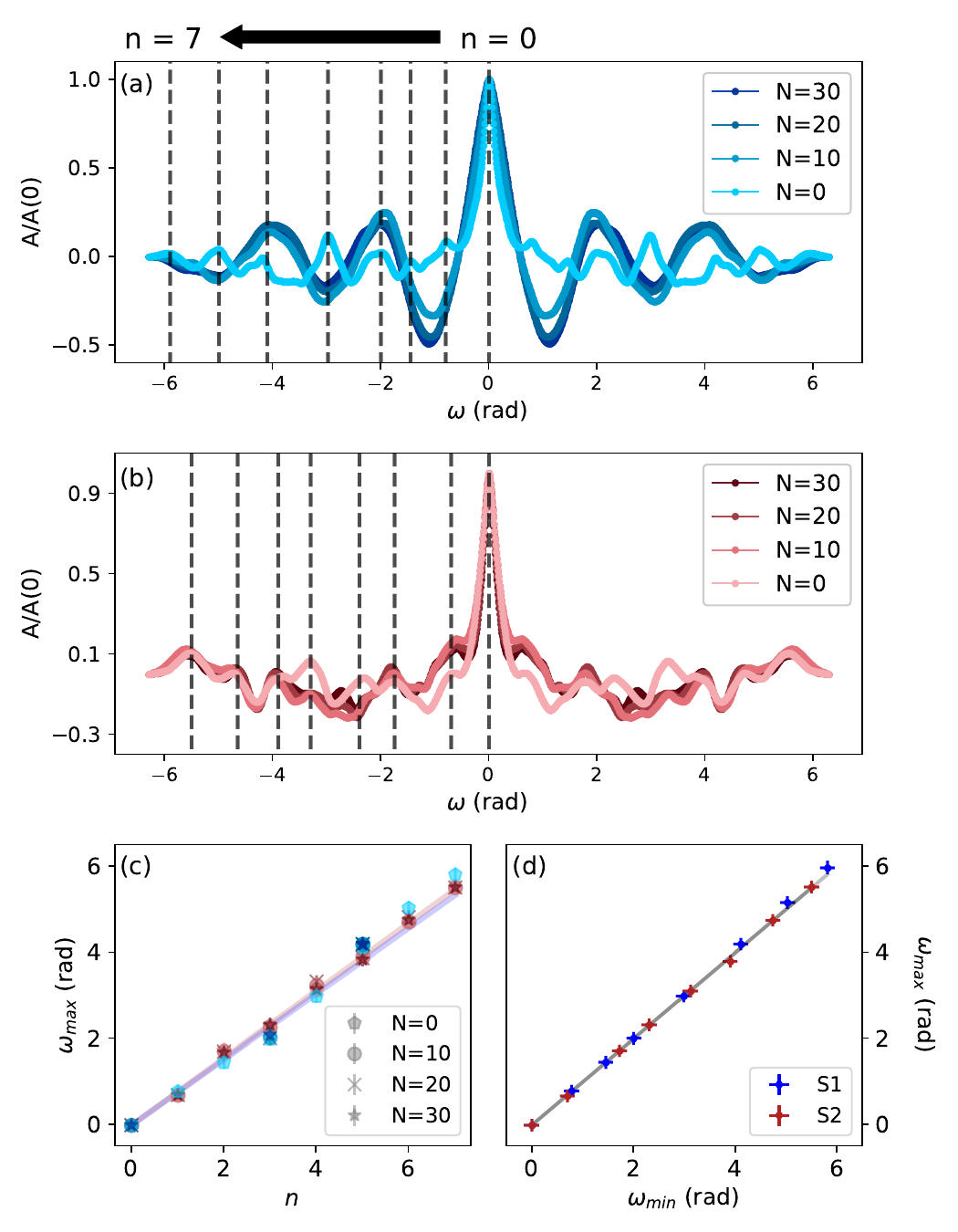}
    \caption{Average $A/A(0)$ curves corresponding to domains evolved after applying $N$ ac square pulses, with $N = 0$, $10$, $20$, and $30$, for samples S1 (a) and S2 (b). Dotted lines indicate the angular positions   of the secondary maxima ($\omega _{max}$) labeled as $n = 1,2,..$. (c) $\omega _{max}$ as a function of $n$ after applying $N$ ac pulses (different symbols) for samples S1 and S2 (blue and red symbols). (d) $\omega_{max}$ as a function of the angular minima of the roughness function ($\omega_{min}$) in domains growth in both samples ($N=0$).}
    \label{fig:posicionmax}
\end{figure}

We now focus on the analysis of the evolution of this large-scale periodicity with the number of ac pulses. Because the radius of the DW changes during the evolution \cite{Art-PRB99-Domenichini,Art-PRB103-Domenichini}, we will consider angular positions in order to compare the secondary maxima. Figure \ref{fig:posicionmax} shows examples of $A/A(0)$ for S1 (a) y S2 (b), as a function of the angular distance $\omega$ for different numbers of ac pulses $N$. We define an index number ($n$) that labels each secondary maximum in increasing order relative to the central maximum of $A(\ell)$ in the initial condition ($N=0$). This labeling is preserved during the evolution $N > 0$, even in cases where a maximum disappears. Figure \ref{fig:posicionmax}(c) show the relative angular positions of the secondary maxima $\omega_{max}$ for each sample as a function of the index $n$ for a different number of ac pulses $N$. After a few pulses, some of the maxima initially present disappear in S1, while in S2 the total number of maxima is preserved during the whole evolution up to $N=30$. Remarkably, as shown in Figure \ref{fig:posicionmax}(c) all the data can be fitted with a unique linear regression with slope $(0.76 \pm 0.02)$ rad for S1 and $(0.78 \pm 0.02)$ rad for S2, which indicates that, in both cases, there is a periodic structure in $u(z)$, and this periodicity is not modified during the evolution. The spatial scale associated with this periodicity ($\sim 30 \, \mu$m in the initial domain size) is large compared with the high-frequency fluctuations but still a fraction ($\sim 1/8$) of the total DW length, and therefore it is naturally associated with the large scale deformations. 

As it was mentioned before, a close relationship between $A(\ell)$ and $B(r)$ is expected. At large distances, where low-frequency deformations become significant and the proposed model for $S(q)$ no longer holds, $B(r)$ also exhibits structured patterns. In fact, by expanding the squared power in Equation \ref{eq:Rug}, it can be shown that $B(r)$ behaves as $-2 <u(z+r)u(z)> $, mining that minima in $B(r)$ may correspond to maxima in the autocorrelation. Figure \ref{fig:posicionmax}(d) shows the linear correspondence with slope $1$ holding between $\omega_{max}$ and the angular position of the minima in $B(r/\rho)$, $\omega_{min}$, at $N = 0$, for samples S1 and S2 respectively. 

We also noticed that, in both samples, the amplitude of the oscillations increases with $N$, in consistence with the large domain deformations observed during the ac evolution \cite{Art-PRB99-Domenichini} evidenced in the increase of $ \langle u ^2\rangle$ (Figure \ref{fig:lav}). 

By analyzing the autocorrelation function at different magnetic fields, we conclude that the positions of the secondary maxima are also independent of the applied magnetic field. The angular period corresponds to linear distances between $28$-$31 \mu$m. We then conjecture that this periodicity is determined by the boundary conditions and some characteristic intrinsic length, probably associated with the typical size of the disorder inhomogeneity, rather than for the driving process. Moreover, this periodicity is larger but on the order of $l_{av}$. This could indicate that the size of the avalanches is limited by the typical size of the disorder inhomogeneity and determines in the same way the DW large-scale structure.

\subsubsection{Small scale fluctuations}

\begin{figure}[h]
  \centering
    \includegraphics[scale=0.5]{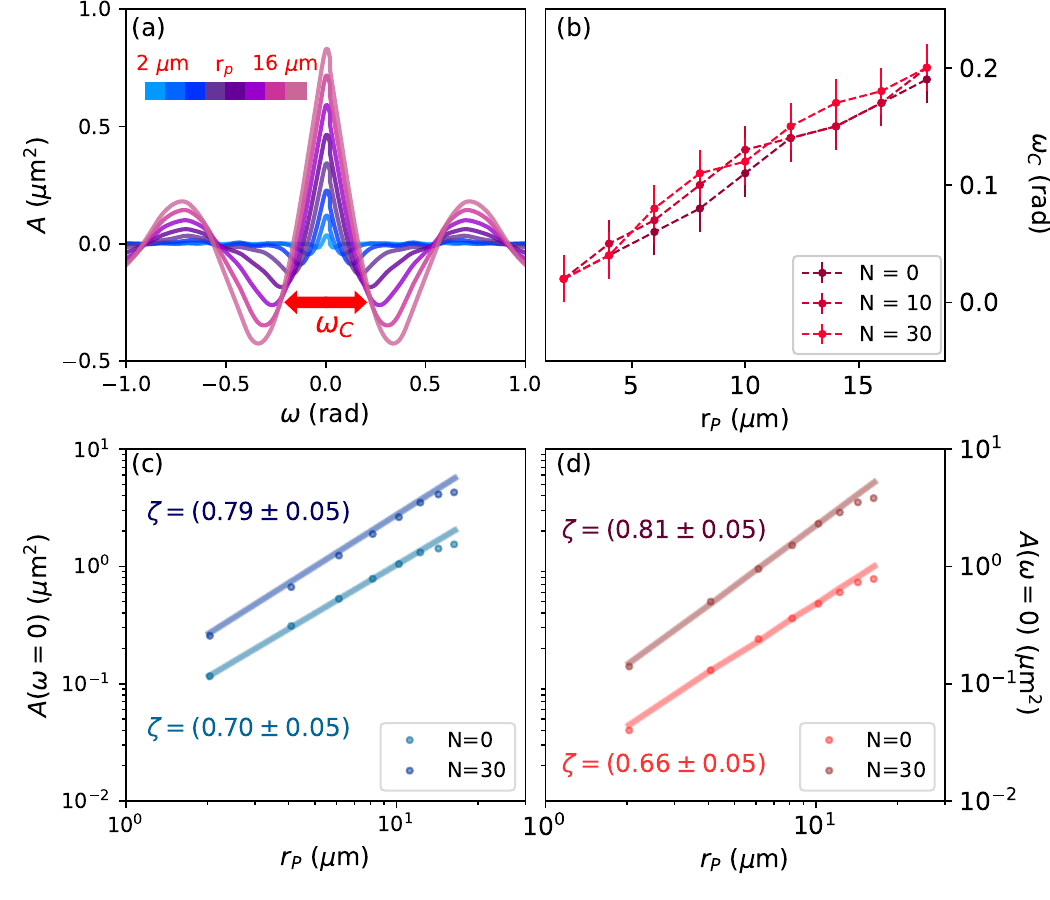}
    \caption{(a) Example of autocorrelations of high-frequency profiles for different values of $r _P$ between 2 $\mu$m and 16 $\mu$m, obtained from a domain grown in the sample S2 with N=0. (b) $\omega _C$ values as a function of $r_P$ for a domain of sample S2 after applying N=0, 10 and 30 ac pulses. Maximum of the autocorrelation function, $A(0)$, for the high-frequency profiles as a function of the averaging distance for $N = 0$ and $N = 30$ in samples S1 (c) and S2 (d).}
    \label{fig:Fig-Perf2}
\end{figure}

Figure \ref{fig:Fig-Perf2}(a) shows examples of the central peak in the autocorrelation obtained for high-frequency fluctuations using window sizes ranging from $r_P = 2 \mu$m to $16 \mu$m (equivalent at $P = 5$ to $40$). It can be seen that, in this case,  both the width and the maximum amplitude increase with the size of the MASF window. Moreover, the width of the peak  increases linearly with the MASF window size and it is independent of the number of ac pulses (Figure \ref{fig:Fig-Perf2}(b)). The lack of any particular characteristic length is in agreement with what is expected in a self-affine interface. 
 
From Expressions \ref{eq:ularge} and \ref{eq:usmall}, together with the standard definition of the mean roughness $\langle u^2\rangle$, it can be easily shown that the high-frequency fluctuations correlation, $\langle u_h^r(z +\ell) u_h^r(z)\rangle$, is equivalent to the mean roughness of a segment of length $r$, as long as $u_h^r(z)$ and $u_l^r(z)$ are uncorrelated and $r<<L$. On the other hand, we know that $A(0)=\langle u_h^r(z)^2 \rangle$ that, in an Edward Wilkinson interface, should present a power law $r^{2\zeta}$. Figure \ref{fig:Fig-Perf2}(c) shows examples of $A(0)$ as a function of $r_P$, for $N=0 $ and after applying $N=30 $ ac pulses on samples S1 (c) and S2 (d). A linear tendency in logarithmic scale is observed for both samples with slopes corresponding to $\zeta$ values between 0.65 and 0.8, in agreement with the effective roughness exponents computed from $B(r)$. The linear fit fails for $r_P>10 \mu m$; this limit is approximately coincident with the limit for the linearity interval considered for $B(r)$ and a significant fraction of the correlation length. Then by analyzing the range of linearity in $A(0)$ in terms of the window size, it is possible to determine the limit between the length range where the power law holds in a simpler and faster way than studying the Pearson´s correlation coefficient to properly fit $B(r)$. 

\section{ Conclusions}\label{sec:Conclusions}

In this work we analyze the roughness of magnetic ``bubble-like" domain walls present in ferromagnetic thin films with PMA, focusing on their initial evolution when applying ac magnetic pulses that strongly deform the domains. We study the evolution of the scaling roughness exponent, holding in the small length scale, as well as that of the statistical pattern of large-scale deformations.
The roughness exponent $\zeta$ is obtained by fitting the roughness correlation function $B(r)$ for $r$ much smaller than the domain size,  where the approximation of self-affinity is valid. By assuming that this $\zeta$ value is an effective exponent associated with various dynamics regimes holding at different length scales, we can estimate the lengths involved in the thermal activation process ($l_{opt}$) and DW avalanches ($l_{av}$). Whereas the first was estimated from parameters obtained from experimental $v(H)$ curves (depinning magnetic field and temperature), the second was obtained from the effective roughness exponent using the model proposed for the structure factor $S(q)$ in \cite{Art-PRB-104-Albornoz}. Values for $l_{opt}$ range between $0.24$ and $0.30$ $\mu$m (smaller than our resolution) whereas $l_{av}$ values range from $5$ to $15 \mu$m in both samples. Both $S(q)$ and $B(r)$ curves obtained from the model with the estimated parameters are in good agreement with the experimental data for large $q$ (small $r$) values. Because under this model the effective $\zeta$ and $l_{av}$ are connected, both evolve similarly under the application of ac magnetic pulses, displaying a subtle increase in the first ac cycles, and then tending to an almost constant value up to $N=30$ cycles. 

We also study the evolution of the autocorrelation function of $u(z)$, an additional useful tool to characterize correlations and characteristic lengths of the DW interface, whose results are in strong connection with those obtained in the previous analysis. The obtained correlation length for large scale (low frequency) fluctuations is very similar to the distance beyond which the proposed $S(q)$ model is no longer valid. By applying MASF windows of different sizes, we separate large-scale (low frequency) and small-scale (high frequency) DW fluctuations. The maximum of the autocorrelation of high-frequency fluctuations scales with the size of the windows with roughness exponents consistent with those obtained from $B(r)$. 

As an interesting and novel result, we show that large-scale deformations, whose amplitude notably increases when applying ac magnetic pulses, display a statistical periodicity: The angular period of the secondary maxima in the average autocorrelation (coincident with those of the minima in the average $B(r)$ at large $r$) does not change during the ac evolution and is independent on the amplitude of the applied magnetic fields. We then propose that this large-scale periodicity is determined by the boundary conditions and some characteristic intrinsic length, probably associated with the typical size of the disorder inhomogeneity. However, the ac driving process indeed affects the amplitude of these oscillations, as well as the evolution of the domain size. The fact that the angular periodicity corresponds to linear distances in the same order that $l_{av}$, could be an indication that, in these bubble domains, the size of the DW avalanches is limited by the typical size of disorder inhomogeneity that determines the DW large scale structure. 

In summary, we are presenting a complete analysis of the domain wall structure in bubbles magnetic domains at small and large length scales, and its initial evolution when applying ac magnetic pulses, a procedure that strongly deforms the domains. The analysis has been done by two alternative methods giving consistent results, both in the critical exponents and characteristic lengths. While the information contained in the autocorrelation function is also present in the commonly used roughness correlation function, from a practical point of view, the first turns out to be particularly useful for determining both the roughness exponent and large-scale deformations. We are confident that the present work would be useful, not only for the community working on the specific topic of magnetic DW, but also for other research areas, involving physical systems able to be modeled as elastic interfaces of finite size in disordered media.

\appendix

\section{ \label{AppendBrSq} Discrete Expressions of $B(r)$ and $S(q)$ and their connection}

For a DW of length $L$ sampled with $N_P$ data points, the roughness function $B(r)$ can be discretely written as

\begin{equation}
    B(r_k)=\frac{1}{N_p-k}\sum_{j=1}^{N_p-k} (u_{j+k}-u_j)^2
    \label{eq:BrYSq}
\end{equation}
\noindent where $r_k=k \delta r$ with $\delta r$ the distance between consecutive points. Assuming $N>>k$ so that $(N-k) \approx N $, and replacing $u _j$ with its Expression in term of its corresponding Fourier transform $\tilde{u}$, the roughness function can be written as

\begin{eqnarray}
B(r_k) = \sum_{j=1} ^{N} ( \sum _{n=1} ^N \sum_{m=1} ^N ( \tilde{u} _n \tilde {u} _m ^* \mathrm{exp} [ i \left ( q_n - q_m \right ) \\
\left ( x_j + r _k \right ) ] + \tilde{u} _n \tilde {u} _m ^* \mathrm{exp} \left [ i \left ( q_n - q_m \right ) x_j \right] + \nonumber \\
+ \tilde{u} _n \tilde {u} _m ^* \mathrm{exp} \left [ i \left ( q_n - q_m \right ) x_j \right] \mathrm{exp} \left [ i q_n r _k \right] + \nonumber \\
+ \tilde{u} _n \tilde {u} _m ^* \mathrm{exp} \left [ i \left ( q_n - q_m \right ) x_j \right] \mathrm{exp} \left [ - i q_n r _k \right]  ) \frac{1}{N}.
\end{eqnarray}

Using the Kronecker delta $\delta_{nm} = \frac{1}{N} \sum_{j = 1} ^N$exp$[i(q _n - q _m) x _j]$, and considering the symmetry of the structure factor \cite{Art-PRB-104-Albornoz,Libro-FeroMag-Guyonnet}, it is possible to write the above equation as the expression shown in Equation \ref{eq:BS}, used to obtain the $B(r)$ function from the $S(q)$ model (blue curves in Figure \ref{fig:Fig-BS} are examples).

\begin{figure}[h]
  \centering
    \includegraphics[scale=0.58]{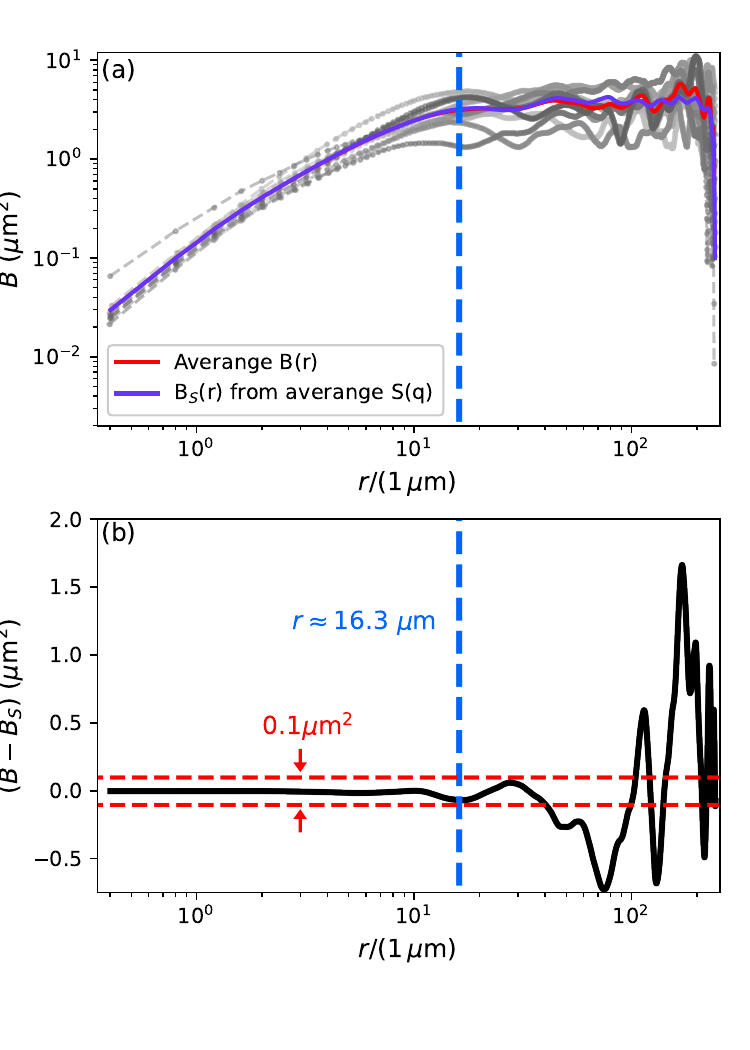}
    \caption{ (a) Roughness correlation function obtained from 10 realizations of a domain grown in nearly identical conditions (gray lines), their average curve (red line), and the roughness function $B(r_k)$ obtained from Equation \ref{eq:Su} using the average structure factor ($B_S$). (b) Difference between the average roughness function $B$ and $B _S$. With a tolerance of $0.1 \mu$m$^2$ (dashed red lines), a validity range of $r < 16.3 \mu$m (blue line) is obtained.}
    \label{fig:Fig-BrDif}
\end{figure}

To estimate the range of $r$ where Equation \ref{eq:BS} is valid for our specific case, we calculated the roughness function by applying this equation to the average experimental structure factor obtained using Equation \ref{eq:Su} ($B_S (r_k)$). Next, the average $B(r_k)$ obtained from the DW profiles using the Equation \ref{eq:BrYSq} was subtracted from $B_S(r_k)$. Figure \ref{fig:Fig-BrDif} illustrates an example of the difference between both functions, which was obtained from domain profiles in sample S2. By applying the criterion that the tolerance range of the subtraction $(B - B_S)(r_k)$ should be less than half of the standard deviation obtained during the calculation of the average roughness function (which, in this particular case, is $0.1 \, \mu m^2$), we determine a validity range of $r < 16.3 \, \mu m$.

\section{ \label{AppendModelSq} Effective Roughness exponent and characteristic lengths}

As mentioned in section \ref{sec:StrucFact}, according to the model proposed in Ref. \cite{Art-PRB-104-Albornoz}, the structure factor (Equation \ref{eq:modeloS}) is expressed in terms of the roughness exponents $\zeta _{eq}$, $\zeta _{dep}$ and $\zeta _{th}$ (corresponding to three stationary states equilibrium, depinning and flow), the amplitude $S_0$, and the characteristic lengths limiting the regions where each of those states will preponderate, $l_{av} = 2 \pi/ q_{av}$ and $l_{opt} = 2 \pi/ q_{opt}$. The free parameters of the model are then $S_0$, $l_{opt}$, and $l_{av}$. 

To determine the optimal parameter set, one approach is to calculate $B(r)$ from Equation \ref{eq:BS} by substituting the proposed $S(q)$ from Equation \ref{eq:modeloS} with different parameter values. These calculated $B(r)$ values can then be compared with the experimental results. However, because the experimental $\ln(B(r))$ is fitted with a linear regression, only two parameters ($B_0$ and the effective $\zeta$) can be determined (see Expression \ref{eq:Rug-exponent}), while $S(q)$ depends on three ($S_{o}$, $q_{av}$, $q_{opt}$). Therefore, this comparison cannot be optimized unequivocally and it is necessary to estimate one of these parameters independently.

The characteristic length $l_{opt}$ is a good candidate for this estimation. From a phenomenological point of view \cite{Art-PRB-104-Albornoz, Art-PRB98-Jeudy} 

\begin{equation}
l_{opt}= L_C \left (\frac{H}{H_d}\right )^{- \nu_{eq}},
    \label{eq:lopt}
\end{equation}

\noindent where $\nu _{eq}=1/(2-\zeta_{eq}) = 3/4$ , and

\begin{equation}
L_C= \left ( \frac{\sigma k_B T_d}{4 M_S^2 t H_d^2} \right ) ^{1/3},
    \label{eq:larkin}
\end{equation}
is the Larkin distance \cite{lemerle1998domain}, being $H_{d}={\sigma \lambda}/{2 M_S L_C^2}$ the depinning magnetic field defined in terms of the micromagnetic parameters $\sigma$ (elastic energy per unit of area) and $M_S$ (saturation magnetization) together with $\lambda$, the characteristic displacement when moving a DW segment of length $L_C$.

Estimations for $H_d$ and $T_d$ can be obtained experimentally by fitting the velocity $v$ as a function of $H^{1/4}$ using Equation \ref{eq:vh} at a fixed known temperature. From this dependence, a linear relationship $\ln (v)= \beta + \alpha (H^{1/4})$, is expected, being $\alpha = (T_d H_d ^{1/4}) / T$ and $\beta = ln(v_d) + (T_d / T)$, where $v$ and $v_d$ are adimensional velocities, expressed in units of m/s. Therefore, the depinning field and temperature can be estimated from the slope $\alpha$ and the intercept $\beta$ of the linear regression as:

\begin{equation}
  T _d = \left ( \beta - \ln v_d \right )  T   ;    H _d = \left ( \frac{\alpha T}{T _d} \right ) ^4.
\end{equation}

On the other hand, $M_S$ and $\sigma$ are known micromagnetic parameters \cite{Art-PRB99-Domenichini}. 
The first rows in Table \ref{Tab:Parameters} show experimental values of $\alpha$ and $\beta$ obtained in both samples at room temperature (see Ref. \cite{Art-PRB99-Domenichini} and \cite{Art-PRB103-Domenichini}), whereas the values of $M_S$ and $\sigma$ was extracted from \cite{Art-PRL99-Metaxas} (S1) and \cite{Art-APL108-Rojas} (S2). It can be seen that $v _d$ takes values between $2$\ and $100$ m/s \cite{Art-APL112-Quinteros}, so $\ln v_d$ is much smaller than $\beta$ and does not substantially modify the estimation of $T_d$.

Last columns in Table \ref{Tab:Parameters} contain the estimated values of $H_d$ and $T_d$ and the Larkin length $L_C$, obtained using \ref{eq:larkin}. 

\begin{table}[H]
\begin{center}
\begin{tabular}{|c|c|c|c|}
\hline
Sample   & Unity     & S1         & S2          \\ \hline
$\alpha$ & Oe$^{-4}$ & 109 (3)    & 164 (3)     \\ \hline
$\beta$  & -         & 25 (1)     & 38 (1)      \\ \hline
$M_S$    & kA/m      & 910        & 540         \\ \hline
$\sigma$ & mJ/m$^2$  & 2.26       & 2.26        \\ \hline
$H _d$   & Oe        & 600 (200)  & 830 (100)   \\ \hline
$T _d$   & K         & 6630 (580) & 10249 (830) \\ \hline
$L _{C}$ & nm        & 80 (30)    & 90 (30)     \\ \hline
\end{tabular}
\caption{Estimated parameters obtained from experimental $v(H)$ and Refs. \cite{Art-PRL99-Metaxas} (S1) and \cite{Art-APL108-Rojas} (S2) for both samples.}
\label{Tab:Parameters}
\end{center}
\end{table}

Finally, from Expression \ref{eq:lopt}, we can obtain $l_{opt}$ for different fields $H$. In the range of fields used in this work, $l_{opt-S1} \simeq 0.30 \pm 0.01 \, \mu$m and $l_{opt-S2} \simeq (0.24 \pm 0.01) \mu$m, smaller than our resolution. 

\begin{figure}[h]
  \centering
    \includegraphics[scale=0.55]{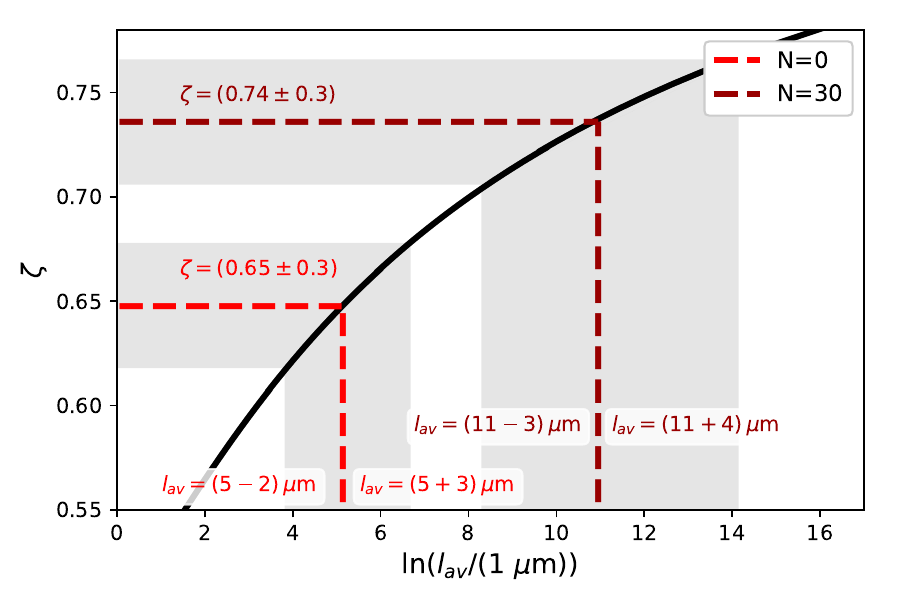}
    \caption{Effective roughness exponent $\zeta$ as a function of the avalanche length $l_{av}$ following the proposed model and the estimated $l_{opt}$ for sample S2 (black line). Horizontal dashed lines are examples of experimental $\zeta$ values obtained in sample S2, in a pristine domain ($N = 0$) and after applying $N = 30$ ac pulses with amplitude $H = 150$ Oe and period $\tau = 50$ms. The procedure used to obtain $l_{av}$ from the experimental values of $\zeta (N)$ is illustrated with vertical dashed lines.   It can be seen that this method yields asymmetric error bars for $l_{av}$ (shaded area).}
    \label{fig:Fig-CurvaExpLav}
\end{figure}

Having obtained an estimation of $q_{opt}=2 \pi/l_{opt}$, $S(q)$ curves for different values of $q_{av}=2 \pi/l_{av}$ were constructed following the proposed model (Equations \ref{eq:modeloS}, \ref{eq:Sj} and \ref{eq:Sth}). The corresponding $B(r)$ was fitted (in the linear log-log region with the same criterion followed in the experimental case), in order to obtain the effective roughness exponent $\zeta (\l_{av})$.
Figure \ref{fig:Fig-CurvaExpLav} shows the function obtained in sample S2 and its interception with several experimental roughness exponents obtained after applying $N$ ac cycles of amplitude $H=150$ Oe. The corresponding $l_{av} (N)$ values were obtained as illustrated in the figure.

\begin{acknowledgements}
We especially thank Javier Curiale, Sebastián Bustingorry, and Alejandro Kolton for their helpful discussions, as well as Mara Granada and J.-M. George for providing the samples. This work was partially supported by Consejo Nacional de Investigaciones Científicas y Técnicas - Argentina and the University of Buenos Aires-Argentina.
\end{acknowledgements}


\vspace{20pt} 

\end{document}